\documentclass{llncs}
\usepackage{llncsdoc}
\usepackage{graphicx}
\begin{document}
\title{Influence of Context on Decision Making \\ during Requirements Elicitation}
\author{Corentin BURNAY , Ivan JURETA \and St\'ephane FAULKNER}
\institute{PReCISE Research Center \& Department of Business Administration \\ University of Namur}
\maketitle

\begin{abstract}
Requirements engineers should strive to get a better insight into decision making processes. During elicitation of requirements, decision making influences how stakeholders communicate with engineers, thereby affecting the engineers' understanding of requirements for the future information system. Empirical studies issued from Artificial Intelligence offer an adequate groundwork to understand how decision making is influenced by some particular contextual factors. However, no research has gone into the validation of such empirical studies in the process of collecting needs of the future system's users. As an answer, the paper empirically studies factors, initially identified by AI literature, that influence decision making and communication during requirements elicitation. We argue that the context's structure of the decision should be considered as a cornerstone to adequately study how stakeholders decide to communicate or not a requirement. The paper proposes a context framework to categorize former factors into specific families, and support the engineers during the elicitation process.
\end{abstract}

\section{Introduction}
The process of analyzing the goals of a system to be and the needs of its future users is commonly referred to as Requirements Engineering (RE). Its first step is requirements elicitation, which consists in collecting information about the expectations and needs of stakeholders, in order to identify the problems that need to be solved by the future information system \cite{roadmap}.

Collecting requirements implies extensive communication between the requirements engineers and the stakeholders. In this communication process, engineers expect stakeholders to state their requirements on the future system. When communicating requirements, stakeholders choose requirements to state on the basis of implicit assumptions on, e.g., the conditions that will hold in the future, when the information system will be operational. The implicit assumptions remain obscure to the requirements engineer, potentially leading to incomplete and conflicting requirements. It is therefore relevant for RE to study how stakeholders decide which requirements to communicate, and what implicit assumptions they make when doing so.

Classical reasoning theories offer a relevant groundwork for this purpose. Their mathematical approach of reasoning makes it easier to relate it to requirements engineering's formalisms. Some of these theories are based on classical logic, which we consider inadequate to relevently model decision making. As an alternative, we focus on non-monotonic reasoning (NMR) theories, which enable a conclusion to be withdrawn after new information invalidating the decision appears. NMR is similar in many ways to a stakeholder's decision making. Indeed, background knowledge in a decision situation is hardly ever complete. NMR consequently assumes that the stakeholder makes normality assumptions when drawing conclusions: anything the stakeholder does not know is assumed to be as expected \cite{Brewka}. Yet, it may happen these assumptions turn out to be incorrect, which may in turn invalidate the conclusions drawn from the assumptions. Such reasoning is called common-sense reasoning, and is considered to be non-monotonic. Reiter \cite{Reiter} proposed default logic, an influential model of non-monotonic reasoning. In default logic, the normality assumption states that, in absence of evidence to the contrary, default rules hold. 

We assume that default logic can relevantly model reasoning that occurs in stakeholder's decision making during requirements elicitation. For example, concluding from the sentences ``The stakeholder wants her accounting software to have the feature to import data from the software which keeps track of wages levels'' that ``The software can actually import data from wages software'' involves for a stakeholder to consider a default assumption such as ``The two software programs can be connected and exchange data''. This is a typical example of a default assumption made by the stakeholder, that is not explicitly communicated to engineers but which constitutes an important requirement. In fact, the default assumption - or the default requirement - may be untrue and invalidate the initial requirement, e.g. the two software programs are proprietary applications closed to connection with other software. During elicitation, the engineer should know what default requirements underline a communicated requirement, since knowing the default requirement helps to decide if the basis requirement that the stakeholder communicated is correctly defined. 

Identifying the default assumptions is not a simple task. There exist many NMR studies that demonstrate the impact of contextual factors on default reasoning \cite{Ford2000,Ford2005,Vogel,HewsonVogel,Elio1,Elio2}. Therefore, engineers should consider that many parameters play a role in the process of adopting or not a default requirement.  In other words, there is a variability in decision making which is inherent to the context. Such variability is likely to lead to some bias between real stakeholders' requirements and what is stated in the result of the elicitation process. We argue the use of a context framework - a list of the different categories of information that forms context - is of great help to identify default requirements of the stakeholders under such circumstances, i.e. information that they implicitly assume in a given context, but do not say explicitly. The requirements engineers could use the framework as a tool to determine which questions to ask during the elicitation in order to identify the default requirements, thereby verifying the completeness and precision of requirements that the stakeholders have provided.

The paper proposes a preliminary discussion and study of factors identified in AI and applied to the context of Requirements Engineering. The paper also reports results from preliminary experiments that suggest the influence of context on requirements elicitation and confirm the need for a new experimental framework to consider different factors and their influence on decision making. The paper then proposes a context framework to categorize factors that influence communication during the requirements elicitation step. 

\section{Internal and External Domains}
We see two distinct categories of factors that are likely to influence stakeholder's decision making. This distinction is important to the requirements engineer, because the two categories do not have the same impact for requirements identification. The engineers should consider such distinction during the elicitation.

The first category is on factors related to human cognition and factors that depend on the way an individual uses knowledge and heuristics in reasoning. It is typically this kind of factors that have been studied by NMR community \cite{Ford2000,Ford2005,Vogel,HewsonVogel}. These are referred to as \textit{internal factors}, since they depend on the individual, and not on the situation in which the individual draws conclusions. Consequently, their implications for the requirements engineers is limited.

The second category includes factors which are not, strictly speaking, specific to the person. They are consequently referred to as \textit{external factors}. These factors also influence how a stakeholder makes the decision to communicate a requirement. As a requirements engineer, it is interesting to understand such factors' influence, because they apply to any stakeholders. It is important to understand how factors influence the choice of default requirements used by the stakeholder, in order to account for such influence in further requirements treatment. Only a few papers \cite{Elio1,Elio2,ShapiroWason} deal with the influence of external factors on NMR. None of them focuses on the case of elicitation. 

External factors are different from internal factors because they may be relevant for one individual, and irrelevant for another: this observation is particularly important in the scope of the RE process. Requirements engineers aiming to design a new political forecasting system should consider factors that influence the entire set of stakeholders, namely the political candidates. To do so, they should establish a clear distinction between internal and external domains. If they do not consider this distinction, they would collect default requirements related to actual needs of one candidate (due to internal factors), but which do not apply to other candidates. For instance, beliefs a candidate has are an internal factor that is likely to influence her expectations regarding the system-to-be and consequently lead to potential bias in the requirements elicitation: the candidate believes that charisma is a key to be elected, and therefore wants the future forecasting system to account for such element in the forecasting function. Yet charisma is maybe not considered as relevant for a second candidate, thereby invalidating the previous requirement. This issue is due to the internality of beliefs: what is true for one is maybe not true for the other, and the factor is consequently considered as internal. On the other hand, the factor ``Size of Group'' identified by Elio and Pelletier \cite{Elio2} is a relevant factor for any candidate, i.e. politician with largest group of potential voters is likely to be predicted as the winner of elections in a democratic political system. This factor is related to the environment of candidates, and not to their own perception of this environment. ``Size of Group'' is therefore classified as an external factor. Such external factors should be considered by engineers as a real and unbiased source of default requirement.

In short, the elicitation of requirements is based on the decision of stakeholders to communicate or not some requirements to engineers. The decision process can be adequately modeled by default logic, in which anything the stakeholder does not know is assumed to be as expected. Such an assumption is called a default, and we will call it in RE a default requirement. Such defaults are opaque to the engineers, making it necessary to attempt to uncover them during elicitation. We argue this can be done using a definition of context, discussed later in this paper. This definition categorizes external factors, that we consider to be acceptable sources of defaults. As an answer to the limited attention to external factors influencing default reasoning, we report results of our preliminary experiments, with the aim to replicate Elio and Pelletier's research in the scope of managerial decision making and RE. Conclusions of this replication lead to a discussion about the limitations of such empirical approach in RE, and suggest the need for a broader experiment design.

\section{Experiment}
\subsection{Design}
So far, few external factors have been identified in the literature, and none of them have been validated in the particular case of RE. The first step toward understanding of NMR in RE decision making during elicitation is to validate the previously identified external factors in relation to RE decision making problems. To test the influence of these factors, we ask subjects to consider a basic problem and to provide a first conclusion. We then introduce similar problems with additional potentially influencing factors. Questions in the questionnaires are benchmark problems requiring ``basic default reasoning''. A benchmark problem always introduces at least two objects that are supposed to respect a rule, then informs the subject one object does not respect the rule, i.e. there is an exception to the rule, and finally asks the subject whether the remaining object respects that rule. 

In our experiment, problems deal with classification of two objects that are described using one or more default rules. Problems then inform subjects that there is at least one exception to the default rules. Finally, subjects are asked to provide a conclusion about the remaining object. The benchmark answer that is expected is that no exception object for a default rule should have impact on the conclusions drawn about any other object for which that rule applies. The different problems submitted to subjects are highly contextual due to the RE orientation of the experiment. Consequently, exercises are not presented under the typical benchmark form \cite{Lifschitz1989}, but rather under the form of a story. The intrinsic structure of problems are always similar. The goal is to give a plausible organizational context to subjects. Consider the following problem as an example of problems proposed in our questionnaire:

\begin{quote}
``An engineer collects requirements for a system to be used in a factory. The engineer typically considers information about employees of the factory to establish a list of requirements {\scriptsize\emph{(Default Rule)}}. Information about wages {\scriptsize\emph{(Object 1)}} and number of working hours {\scriptsize\emph{(Object 2)}} are examples of information about employees. However, information about wages has not been considered by the engineer {\scriptsize\emph{(Exception)}}. What do you think about engineer's behavior toward working hours? {\scriptsize\emph{(Benchmark Question)}}''
\end{quote}

In the case of the above problem, we expect subjects to select a conclusion from a list of four different answers regarding the remaining working hours information \emph{(Object 2)}:

\begin{enumerate}
	\item Benchmark - \emph{Exception} has no bearing on the \emph{Benchmark Question}: e.g. \emph{Object 2} have been considered by the engineer unlike \emph{Object 1};
	\item Exception - \emph{Exception} also applies to the \emph{Benchmark Question}: e.g. \emph{Object 2} have not been considered by the engineer like \emph{Object 1};
	\item Other - \emph{Exception} does imply another exception, but with different characteristics: e.g. a wage per hour measure have been preferred by the engineer;
	\item Can't Say - The subject cannot choose one of the former proposition.
\end{enumerate}

In order to decrease the chance of finding out the pattern of good answers, the four solutions are randomly ordered, i.e. the benchmark answer will not always be the first answer. Our experiment is designed to confirm the influence of five external factors: Specificity, Similarity, Size of the Group, Nature of the Group, Perspective. The experiment only aims to determine whether already known factors can be confirmed in a RE context. The purpose is not to understand how people are reasoning in terms of the steps that they may be taking,  but what may influence them while reasoning.

\paragraph{Similarity}
refers to the availability of information regarding the similitude between objects. It describes the set of commonalities that are shared by objects, and has two levels, i.e., low or high. For instance, high similarity would inform subjects that \emph{Object 1} and \emph{Object 2} are both provided by the accounting department. The low similarity would state that the two objects come from different departments. Such factor should impact reasoning since high similarity suggests that objects may have been subject to the same set of intentions \cite{Elio2}.

\paragraph{Specificity}
refers to the availability of information about the way the exception is violating the default rule. It is any piece of information that acts as a justification of the exception, and has two levels, i.e., vague or specific. The specific level would state taht \emph{Object 1} is not following \emph{Default Rule} because it is violating privacy rules. Low Specificity level would only explain that \emph{Object 1} was not considered by the engineer, without any explanation about the way the exception happened.

\paragraph{Nature of the Group}
refers to whether the considered entities are real or artifact objects. Wason and Shapiro \cite{ShapiroWason} argue that the difficulty to reason about abstract objects may be due to ``the failure to generate alternatives in order to derive the correct solution''. Facing something she does not think real, a subject will have difficulties to reason about it, and this can impact her final conclusion. The artifact level would state that Object 1 is an artificial one, e.g. a number of winks. Note that we interpret the word artificial as ``out of the plausible set of possibility for a given context''.

\paragraph{Size of the Group}
refers to the relative importance of classes that are compared in the benchmark problem. The factor has two levels: with or without size information. The former would state that only 20\% of wages information is accessible against 95\% of working hours information. The later would avoid any reference to the relative size of objects.

\paragraph{Perspective}
refers to the point of view adopted by subjects. Elio and Pelletier tested subjects answering to questions while taking the ``perspective'' of a human or of a robot. They observe that robots should be cautious \cite{Elio1}, and thereby that people tend to privilege the ``Can't Say answer'' when they are asked to adopt the robot perspective. We replicate the ``who is answering?'' question with different actors more suitable to a RE approach. Our experiment proposes ``For Me'' and ``For Other'' answers.

\subsection{Questionnaires}
In our experiment, we test interactions between two factors. Considering the definition of factors, we focus on the interaction between similarity and specificity on the one hand, and nature and size on the other hand. Perspective is considered for both combination of factors. Questionnaires consist of 21 problems, which are designed to test previous factors. Three questions have been created in order to cover the possible routine of the questionnaire and to make subjects more careful about their way of thinking and taking information into account. These three additional questions are constructed with exactly the same structure as the 18 relevant exercises, but they differ in number of default rules and objects to which rules apply. This makes the reasoning even more difficult, but does not influence the individual in other exercises. The answers to these latter problems will not be taken into account for the final analysis and are designed to avoid bias while entertaining volunteers.

\subsection{Subjects}
The first questionnaire (``For Other'') have been submitted to a group of 68 management science students at the Department of Business Administration, University of Namur. All subjects were bachelor students and were asked to answer within a 50 minutes time period. The second questionnaire (``For Me'') has been submitted to the same group, at a later date. Both sessions took place at the same place and under the same circumstances. Subjects were asked to answer during class time, and were not compensated for participating in the study. The sample is considered to be representative of the population. Firstly, we study how human makes decision in a given context: there should be no difference between decision making performed by students and by engineers, since reasoning is a process that any human performs. Secondly, subjects are management students, trained to make decisions under uncertainty as in real conditions. Thirdly, we test external factors, i.e., factors that are given to subjects and have the same influence for students and for requirements engineers. Consequently, we consider the external validity of this experiment as acceptable.
 
\subsection{Procedure}
Questions were distributed into two questionnaires: we refer to them as ``For Me'' and ``For Other''. Once divided, all the questions were randomly selected and inserted in their own questionnaire. The answerer can use any kind of material. The assignment clearly mentions that there is no best answer, but that some answers are better than others. It also tells the subject that the objective of the questionnaire is to better understand how managers are reasoning in a situation of general and imperfect information about a decision problem.

\subsection{Results}
Table \ref{Table_Point_of_View_SP} provides observed proportions of answers for each point of view. It appears that the perspective adopted by subjects when they answer questions has an impact on the decision they make. Answering for another person seems to make subjects more reluctant to select the benchmark. Rather than selecting an inappropriate answer, subjects prefer the ``Can't Say'' answer. This could be interpreted as cautiousness: because others will suffer from potential drawback related to subjects' decision, they prefer not to give an answer at all.

\begin{table}
\begin{center}
\caption{Proportion of Answers by Perspective}
\begin{tabular}{lllll}
\hline \noalign{\smallskip} 
& Benchmark~~ & Exception~~ & Other~~ & Can't Say~~\\
\noalign{\smallskip}
\hline
\noalign{\smallskip}
For Other~~ & .467 & .079 & .165 & .289 \\ 
For Me~~ & .579 & .151 & .083 & .187 \\
\hline 
\end{tabular}
\label{Table_Point_of_View_SP}
\end{center}
\end{table} 

Significance tests are performed using the same approach as Elio and Pelletier, namely repeated measure ANOVA on the proportion of answers in each category of answer and for each problem \cite{Elio1}. Elio and Pelletier's approach implies that the category of answer is another factor. Perspective's influence is confirmed through a significance test. A significant effect is observed for the answer category. Results also suggest a significant interaction between answer category and Perspective [F(3, 183)=5.923, p=0.000].

Beside Perspective, other factors are tested in this experiment. Table \ref{Answers} summarizes proportions of answers for each possible combination of factor, for both perspective. The reference question is considered as the Vague-Low and Real-Without Number versions of problems. In a nutshell, these are problems where factors' modality are neutral.

\begin{table}
\begin{center}
\caption{Proportions of Answers}
\begin{tabular}{lllll}
\hline \noalign{\smallskip} 
& Benchmark~~ & Exception~~ & Other~~ & Can't Say~~ \\
\noalign{\smallskip}
\hline 
\noalign{\smallskip} 
Specific-High & .559 & .088 & .191 & .162  \\ 
Specific-Low & .485 & .059 & .206 & .250 \\ 
Vague-High & .338 & .221 & .221 & .221 \\ 
\hline 
\textbf{Reference ``For Others''} & \textbf{.500}	& \textbf{.059}	& \textbf{.162} & \textbf{.279} \\ 
\hline 
Artefact-With & .485 & .059 & .147 & .309 \\ 
Artefact-Without~~ & .456 & .044 & .147 & .353 \\ 
Real-With & .412 & .044 & .103 & .441 \\
\noalign{\smallskip}
\hline \hline
\noalign{\smallskip} 
Specific-High & .613 & .194 & .032 & .161 \\ 
Specific-Low & .742 & .032 & .113 & .113 \\ 
Vague-High & .339 & .452 & .065 & .145\\ 
\hline \textbf{Reference ``For Me''} & \textbf{.645} & \textbf{.081} & \textbf{.081} & \textbf{.194} \\ 
\hline Artefact-With & .565 & .097 & .081 & .258 \\ 
Artefact-Without~~ & .581 & .081 & .065 & .274 \\ 
Real-With & .500 & .194 & .145 & .161 \\ 
\hline
\end{tabular}
\label{Answers}
\end{center}
\end{table} 

The reference question for the ``For Other'' perspective has 50\% of benchmark answer, while it ranges between 50\% and 75\% for the Robot perspective of Elio and Pelletier \cite{Elio1}. Subjects in our experiment do better under the ``For Me'' perspective, with a 65\% part of benchmark answer, against an average below 60\% for Elio and Pelletier. Table \ref{Answers} suggests mitigated results. On the one hand, variations in proportions of answers for the Specificity and Similarity are observed, which make the influence of factors visible. On the other hand, variations for the Nature and Size are minimal.

The same conclusion is drawn for the ``For Me'' perspective. One of the most surprising aspects is probably the majority of ``Exception'' in the case of Vague-High questions. Under the ``For Me'' perspective, influence of Similarity and Specificity is evident. Size and Nature slightly decrease the proportion of benchmark in favor of the ``Can't Say'' answer, but no clear pattern appears.

Results do not give clear conclusions. Impact of factors in the two perspectives are identical, even though influence is stronger under the ``For Me'' perspective. In the case of Similarity and Specificity, impact is clearer, particularly in the case of the ``For Me'' perspective. Under this perspective, a difference of 35\% is observed between the two factors. Tests displayed in Table \ref{p_val} confirm that some external factors identified in the literature review do not have, in our experiment, the effect that they have been argued to have in prior research.

\begin{table}
\begin{center}
\caption{Significance Test for the Factors Dimension}
\begin{tabular}{lllll}
\hline \noalign{\smallskip} & \multicolumn{2}{l}{For Other} & \multicolumn{2}{l}{~~~For Me} \\ 
\noalign{\smallskip} 
\hline
\noalign{\smallskip} Factor(s) & Fisher & P-Value & ~~~Fisher & P-Value\\
\noalign{\smallskip} \hline \noalign{\smallskip}  Answer & 18.486 & .000*** & ~~~44.940 & .000*** \\ 
Answer*Similarity & 1.966 & .120 & ~~~16.273 & .000***\\ 
Answer*Specificity & 2.468 & .063* & ~~~7.725 & .000***\\ 
Answer*Specificity*Similarity~~ & 2.363 & .072* & ~~~3.349 & .020** \\ 
\noalign{\smallskip} \hline \noalign{\smallskip}
Answer & 19.856 & .000*** & ~~~30.457 & .000*** \\ 
Answer*Nature & .328 & .805 & ~~~1.672 & .174\\ 
Answer*Size & .666 & .574 & ~~~1.758 & .157\\ 
Answer*Nature*Size~~ & 2.296 & .079* & ~~~1.420 & .238 \\ 
\hline 
\end{tabular}
\label{p_val}
\end{center}
\end{table} 

For the two Perspectives, there is always a significant influence of answer. Table \ref{p_val} shows that influence of Similarity and Specificity is confirmed under the ``For Me'' perspective. Regarding the ``For Other'' questions, a two-ways interaction between answer category and Specificity [F(3,201)=2.468, p=0.063], together with a three-ways interaction between answer category, Similarity and Specificity [F(3,201)=2.363, p=0.072] is observed. Impact of Similarity in such perspective is not confirmed. Results for the ``For Me'' perspective are negative. P-values are often larger than 15 percent and influence of Size and Nature cannot be concluded. We emphasize here the conspicuous influence of Perspective on factors being tested. A single factor has always a clearer impact under the ``For Me'' perspective. The combination of factors leads sometimes to a strengthening effect [F(3,201)=2.298, p=0.079] under the ``For Other'' perspective.

\subsection{Preliminary Conclusion}
Our experiment confirms the influence of Similarity and Specificity as observed in prior NMR research, in RE context. It also highlights how factors' influence depends on the conditions in which they are tested. Perspective significantly impacts the strength of factors, and factors like Size or Nature seems to have no influence on reasoning unless when working together. Results obtained in this experiment are different in some regards from what Elio and Pelletier propose in their own studies. This can be explained by the way problems are introduced to subjects. The importance of studying factors influencing decision making during the elicitation of requirements cannot be denied. However, we observe that classical empirical approach used in NMR literature is not well suited to tackle the issue of elements influencing decision making during elicitation. Factors cannot be identified and tested in an isolated way. We argue these factors are part of a broader concept, namely the context.

In the second part of the paper, it is argued that a set of factors defines a context, and that this context influences selection of default requirements. The factor itself is only a mean to define the content of a problem. The standalone factor has no clear role without considering the context in which it occurs. Since factors influence context, and context influences decision making, it is important to keep a control on factors. Not only those that are tested, but also those that form context and that are not identified as being in the scope of the experiment. This position justifies why our experiment and the experiment from Elio and Pelletier are so different: contexts are different. Some factors related to RE have probably interacted with external factors initially identified, thereby leading to a significantly different context compared to what Elio and Pelletier originally studied. Since decision making is strongly related to the context, it is interesting to study what conditions or factors influence the decision process.

\section{Context as a Reference for Requirements Elicitation}
Experiments performed on external factors are always based on a context. Sometimes, the content of a context is limited, with only a few aspects being explicit. This is the case of a simple Benchmark, proposed by Lifschitz \cite{Lifschitz1989} and broadly used in NMR research.

\begin{quote}
\begin{center}
\textsl{A and B are heavy.\\
Heavy blocks are normally located on this table. \\
A is not on this table. \\
What about B? \\
B is on this table.}
\end{center}
\end{quote}

 In such case, context is a room with a table and the influence of undefined factors can probably be neglected: no actor interacts, no information can influences the individual. In the case of this paper however, context is more complex due to the richer decision problem. As a consequence, unidentified factors must be considered because they can interact with elements to test, and alter the influence these standalone elements could actually have in a basic context.

As an answer, we propose another approach of external factors. We argue that the ``Pick up and test factor'' method is not adequate because factors are part of a context, and as a consequence should be considered within this context. Doing so requires a relevant definition of context. Given the RE orientation of this paper, we consider definitions of context that are proposed in literature on context-aware computing. Numerous lexical definition of context exist. However, few of them enable to consider how factors relate to context. As a consequence, we focus on three particular definition of context: Dey's general context definition \cite{Dey}, Lenat's decomposition of context into twelve dimensions \cite{Lenat} and Zimmermann et al's de operational definition of context \cite{Zimmer}. Each of these definitions emphasizes different ``categories'' or ``dimensions'' of the context. Table \ref{tab} summarizes context's categories identified by former authors:

\begin{table}
	\centering
		\caption{Categories of Context}
		\begin{tabular} {lccccc} \hline \noalign{\smallskip}
		 & Location / Time & Individuality & Knowledge & Relation & Activity \\ \noalign{\smallskip} \hline \noalign{\smallskip}
		Dey& X & X &  & X &  \\
		Zimmermann et al.& X & X &  & X & X  \\
		Lenat& X & X & X & X &  \\ \noalign{\smallskip} \hline \noalign{\smallskip}
		\end{tabular}
	\label{tab}
\end{table}

``Location/Time'' category consists of every factor that is related to the time when the context occurs or to the place where  it can be located. ``Individuality'' category forms the basis of the context. It gathers all the factors that are related to entities existing inside the context. An entity can be of different nature: natural, human or artificial \cite{Zimmer}. ``Granularity'' consists of all the factors dealing with the quantity or the level of information that is provided about the context (whatever its nature). ``Activity'' refers to factors that deal with the set of goals and intentions of individuals existing in the context. ``Relationship'' corresponds to any factor dealing with the relationships between previously defined entities, i.e. in what way two or more entities are related to one another. ``Knowledge'' refers to the information that is part of the context and not specifically attached to an individual. An example is law: individuals know that laws exist, yet they do not always know their content. Laws are parts of ``external'' knowledge that depends on the environment and not on the individual.  

Based on Table \ref{tab}, we propose a context framework which consists in six major categories : Spatio-Temporal, Items, Knowledge, Relationship and Activity categories. The framework gathers categories of context that were not proposed together in former definitions. Based on this framework, it is possible to capture the major aspects of context that influence communication and decision during RE elicitation. The framework enables to define the different categories that form context and to classify factors that are proven to influence decision making.

The framework could be used in several ways. Firstly, it could be used as a tool to support future empirical research on human reasoning. For instance, it enables to account for the differences between the experiment that is proposed in this paper and experiments from AI literature. The reason why replication of results fails in our experiment is that we actually replicate a single aspect of context, e.g. granularity category with similarity or specificity factors. We did not pay specific attention to the remaining categories of context. Thereby, we designed an experiment with different experiment settings and consequently different impacts of factors on default reasoning. Therefore, the framework could be used to list items - or other categories - in order to accurately identify the context's definition and decreasing the gap in experimental settings.

Secondly, we consider that the framework offers numerous opportunities for requirements engineers. It could support the requirements engineers in the identification of the complete set of factors that are likely to have a bearing on the communication with stakeholders. The advantages of categorizing elements of the context in well defined dimension are multiple. Firstly, it is a way to structure a requirements elicitation process by considering each category of the framework. Secondly, the categorization offers a taxonomy to support further formalizations, by decomposing a blurry context notion into a well structured one. Thirdly, it enables to avoid subjectivity during elicitation, since external elements of the context are given to each stakeholder. Finally, it is possible to relate categories of context to other important RE notions: does that part of the context make sense to the user or the machine? Who or what can sense it? Is it relevant regarding the system-to-be? 

\section{Conclusion}
The paper presents preliminary results of an empirical study on decision making during requirements elicitation. We focus our attention on non-monotonic theories, and more precisely on default logics, which offer an adequate groundwork to empirically study human decision making. The paper discusses the distinction between internal and external factors, and tests factors initially studied by Elio and Pelletier in the case of RE. We find in our experiment different results than what is proposed in the NMR literature. We argue these differences can be explained by the definition of the context on which individuals base their decisions. As a response, we suggest to transfer our empirical efforts on the study of context. Our claim is that during elicitation, the requirements engineer should try to obtain as much information about context as possible, in order to uncover default requirements. By looking empirically at which dimensions of context are relevant, we provide a definition of context useful for the elicitation of requirements. Requirements engineers can use that definition - which lists dimensions of context - as a tool to determine the questions they should ask stakeholders to identify stakeholder's default requirements, and thereby verify the completeness of requirements that the stakeholders have provided. Further research is required to validate the context framework, but we believe the framework can be a useful tool to support requirements engineers during requirements elicitation.

\bibliographystyle{splncs}
\bibliography{bibfile1}

\end{document}